\pdfoutput=1	

\documentclass{elsart3p}
\usepackage{graphicx}
\usepackage{amssymb}

\begin{document}


\newcommand{\rmi}{\mathrm{i}}
\newcommand{\rmd}{\mathrm{d}}
\newcommand{\bi}[1]{\mathbf{#1}}

\begin{frontmatter}

\title{Standard and non-standard metarefraction with confocal lenslet arrays}

\author{Johannes Courtial}

\address{Department of Physics \& Astronomy, Faculty of Physical Sciences, University of Glasgow, Glasgow G12~8QQ, UK}

\ead{j.courtial@physics.gla.ac.uk}

\begin{abstract}
A recent paper demonstrated that two lenslet arrays with focal lengths $f_1$ and $f_2$, separated by $f_1 + f_2$, change the direction of transmitted light rays approximately like the interface between isotropic media with refractive indices $n_1$ and $n_2$, where $n_1 / n_2 = - f_1 / f_2$ [J.~Courtial, New J.\ Phys.\ \textbf{10}, 083033 (2008)].
This is true if light passes through corresponding lenslets, that is lenslets that share an optical axis.
Light can also pass through different combinations of non-corresponding lenslets.
Such light can be either absorbed or allowed to form ``ghost images''; either way, it leads to a limitation of the field of view of confocal lenslet arrays.
This paper describes, qualitatively and quantitatively, a number of such field-of-view limitations.
\end{abstract}



\end{frontmatter}


\section{Introduction}

\begin{figure}
\begin{center} \includegraphics{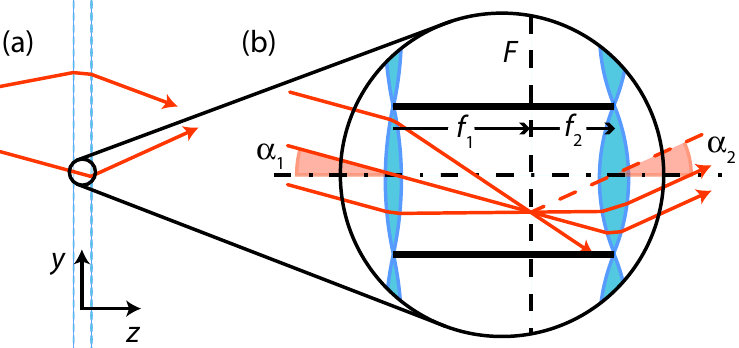} \end{center}
\caption{\label{lens-figure}Light rays passing through confocal-lenslet-array (CLA) sheets.
(a)~The CLA sheet is in the $xy$ plane; the figure shows a $yz$ projection.
(b)~A lenslet in the first (left) array focusses all incident light rays with angle of incidence $\alpha_1$ to a point in the focal plane $F$.
Of those rays, those that pass through the corresponding lenslet in the second (right) array  (that is, the lenslet in the second array with the same optical axis) leave it with an angle of metarefraction $\alpha_2$, where $f_1 \tan \alpha_1 = - f_2 \tan \alpha_2$.
$f_1$ and $f_2$ is the focal length of the lenslets in the first and second array, respectively.
Optional absorbers (thick black horizontal lines) can remove light rays that would otherwise pass through a non-corresponding lenslet in the second array.
(Adapted from Ref.~\cite{Courtial-2008a}.)}
\end{figure}

A recent paper \cite{Courtial-2008a} introduced the idea of using a sheet comprising two lenslet (or microlens) arrays to mimic refraction.
The two lenslet arrays have to share the same focal plane: they are confocal.
The basis of this idea is that the equations describing the light-ray-direction change due to confocal lenslet arrays (CLAs) and refraction due to refractive-index interfaces are very similar:
in the former, the angles $\alpha_1$ and $\alpha_2$ with which a light ray respectively enters and exits a CLA sheet (Fig.\ \ref{lens-figure}) are related through the equation
\begin{equation}
\tan \alpha_1 = \eta \tan \alpha_2,
\label{tan-equation}
\end{equation}
where
\begin{equation}
\eta = -\frac{f_2}{f_1}
\end{equation}
is minus the ratio of the focal lengths of the two lenslet arrays \cite{Courtial-2008a};
in the latter, the angle of incidence in a medium with refractive index $n_1$, $\alpha_1$, and the angle of refraction in a medium with refractive index $n_2$, $\alpha_2$, are related through Snell's law,
\begin{equation}
\sin \alpha_1 = \frac{n_2}{n_1} \sin \alpha_2.
\label{Snell-equation}
\end{equation}
For small angles, when $\sin \alpha_{1,2} \approx \tan \alpha_{1,2}$, the equation describing the light-ray-direction change due to CLAs (Eqn (\ref{tan-equation})) is the same as Snell's law, whereby
\begin{equation}
\eta = \frac{n_2}{n_1}.
\end{equation}

\begin{figure}
\begin{center} \includegraphics{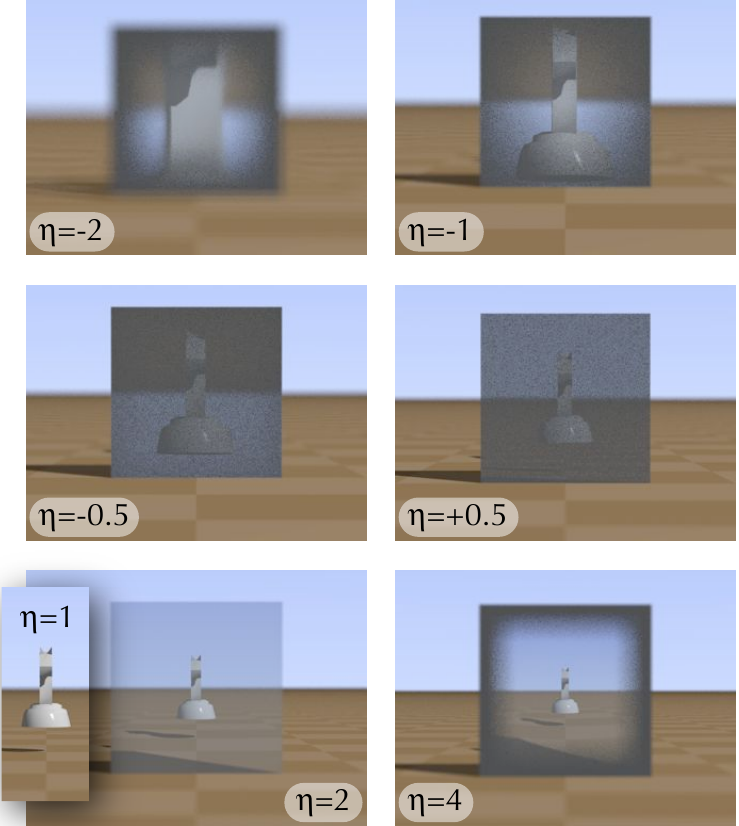} \end{center}
\caption{\label{etas-figure}Chess piece seen through CLA sheets with different focal-length ratios $\eta = -f_2/f_1$ and, for comparison, seen directly (inset labelled $\eta = 1$).
The chess piece is in the same position in all frames, but the longitudinal imaging properties of each sheet makes the chess piece's distance behind the sheet appear stretched by a factor $\eta$.
The brightness of the view is clearly different for different values of $\eta$.
In a few frames, most notably those corresponding to $\eta = -2$, $\eta = -1$ and $\eta = 4$, the brightness even changes across the view.
The reduction in brightness is due to light that would have undergone non-standard metarefraction being filtered out by absorbers (see Fig.\ \ref{lens-figure}(b)).
The frames in this figure are a detailed ray-tracing simulation through the structure of CLA sheets, each comprising $2 \times 200 \times 200$ lenslets, created using the freely-available software POV-ray \cite{POV-Ray}.
The geometry of the lenslet arrays is described in more detail in Ref.\ \cite{Courtial-2008a}.
A movie (MPEG-4, 360 KB) of the view through CLA sheets with the value of $\eta$ changing can be found in the supporting online material.
(Adapted from Ref.~\cite{Courtial-2008a}.)}
\end{figure}

For larger angles, Eqn (\ref{tan-equation}) is different from Snell's law; it does, for example, not lead to total internal reflection.
However, it could be argued that the light-ray-direction change according to Eqn (\ref{tan-equation}) is better than refraction due to Snell's law, in the sense that it has the following remarkable imaging property:
a planar CLA sheet images all space with transverse magnification one and longitudinal magnification $\eta$ \cite{Courtial-2008a}.
(In contrast, a planar refractive-index interface described by Snell's law images all of space only if $n_1 = \pm n_2$; in all other cases, it images only approximately.)
Fig.\ \ref{etas-figure} shows an object seen through CLA sheets for different values of $\eta$, demonstrating the apparently different distance of the object behind the CLA sheets due to the longitudinal magnification,~$\eta$.

CLA sheets are examples of METATOYs (\underline{meta}ma\underline{t}erials f\underline{o}r light ra\underline{y}s) \cite{Hamilton-Courtial-2009}.
Other METATOYs include generalizations of CLAs \cite{Hamilton-Courtial-2009b} and combinations of Dove-prism sheets \cite{Hamilton-Courtial-2008a,Courtial-Nelson-2008,Hamilton-et-al-2009}.
The name stems from a number of similarities with metamaterials \cite{Smith-et-al-2004}, for example structural similarities (both metamaterials and METATOYs are arrays of small elements) and ray-optical negative refraction \cite{Courtial-2008a,Courtial-Nelson-2008}, that is, negative refraction without negative group velocity \cite{Veselago-1968} or amplification of evanescent waves required for the sub-wavelength imaging properties of superlenses \cite{Pendry-2000} and hyperlenses \cite{Liu-et-al-2007,Smolyaninov-et-al-2007}.
Following Ref.\ \cite{Hamilton-Courtial-2009}, the light-ray-direction change due to passage through METATOYs is called \emph{metarefraction}; considering the close similarity between Eqns (\ref{tan-equation}) and (\ref{Snell-equation}), this seems appropriate enough.
$\alpha_1$ is called the angle of incidence and $\alpha_2$ the angle of metarefraction.

The properties of CLA sheets as discussed above apply only to light that passes through corresponding lenslets, that is two lenslets sharing the same optical axis (like the ones in Fig.\ \ref{lens-figure}(b)).
Here this is called \emph{standard metarefraction}.
Light that enters through one lenslet and exits through a lenslet other than the corresponding lenslet is re-directed differently.
This is called \emph{non-standard metarefraction}.

\begin{figure}
\begin{center}
\includegraphics{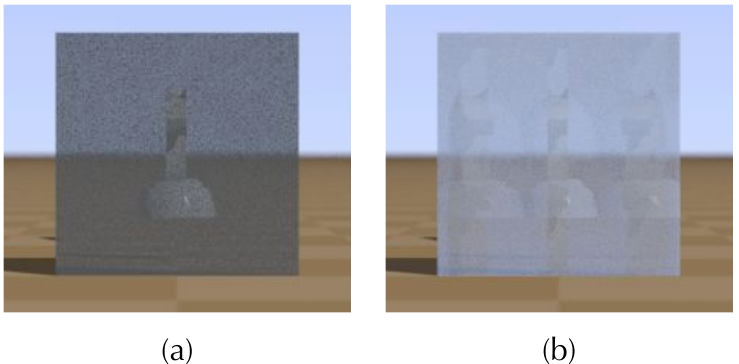}
\end{center}
\caption{\label{ghostImages-figure}Simulated view of a chess piece through CLA sheets with (a) and without (b) absorbers that remove light that would otherwise undergo non-standard metarefraction (see Fig.\ \ref{lens-figure}(b)).
In all other aspects, the CLA sheets are identical.
In the former case, the brightness is reduced; in the latter case, additional images (``ghost images'') of the chess piece appear.
The CLA sheets shown here were calculated for $\eta = 0.5$.
A movie (MPEG-4, 284 KB) of the view through CLA sheets with no absorbers as the value of $\eta$ is varied is contained in the supporting online material.}
\end{figure}

The possibility of non-standard metarefraction was already noticed in Ref.\ \cite{Courtial-2008a}, and in all the simulations in Ref.\ \cite{Courtial-2008a} light that would otherwise have undergone non-standard metarefraction was filtered out with appropriately placed absorbers, leading to a darkening of part of the view (see Fig.\ \ref{etas-figure}). 
If such light is not filtered out, it leads to ``ghost images'': additional images an object seen through a CLA sheet (Fig.~\ref{ghostImages-figure}).

The work on CLA sheets is closely related to ``integral'' photography, a method for taking (and viewing) three-dimensional (3D) photos \cite{Lippmann-1908}.
In integral photography, the second lenslet array views a photo of the intensity distribution created by the first lenslet array, instead of viewing it directly, as in the case of CLA sheets.
Integral photography can also be seen as the basis of lenticular printing, the technique used to create pictures (for example postcards) that provide two or more different images when seen from different angles, or even 3D views~\cite{Anderson-1970}.
It is also the basis of many 3D displays \cite{Okoshi-1980,von-den-Hoff-Flinn-2008}.
A setup that is essentially the same as the camera in integral photography can also be used in a completely different way: instead of aiming for 3D imaging, such ``multiaperture imaging'' \cite{Tanida-et-al-2001} is used in combination with digital processing of the image obtained in the focal plane and concentrates on the small aperture of the individual lenses, which can include faster optics and lower aberrations \cite{Athale-et-al-2008}.
The digital processing can also lead to superresolution \cite{Kanaev-et-al-2007}.
In these contexts, the field of view of lenslet arrays has been researched; a good review of methods to deal with ``parasitic images'' (the equivalent of what is called here ``ghost images'') and corrections to other lens aberrations can be found in Ref.~\cite{Daniell-2005}.

The CLAs discussed here, in which the focal lengths of the two lenslet arrays are different \cite{Courtial-2008a}, are a generalization of basic CLAs in which both lenslet arrays have the same focal length.
The latter have been used in pseudoscopic (depth-inverting) imaging systems \cite{Davies-McCormick-1993,Stevens-Harvey-2002,Okano-Arai-2002} (in the case of Ref.\ \cite{Okano-Arai-2002} in the form of arrays of graded-index lenses).
Applications include correcting the pseudoscopic images provided by integral photography~\cite{Stevens-Harvey-2002}.


This paper is aimed at providing a more detailed analysis of the conditions under which standard and non-standard metarefraction occur in CLAs in which the lenslet arrays have different focal lengths, and how non-standard metarefraction manifests itself visually.
This leads to a qualitative and quantitative understanding of the field-of-view limitation in such CLA sheets.


\section{\label{metarefractions-section}Standard and non-standard metarefraction for different angles of incidence and metarefraction}

In this section light-ray propagation through one particular pair of corresponding lenslets in CLA sheets is considered.
The argument assumes that the lenslets have the same, circular, aperture and that corresponding lenslets share the same optical axis.

\subsection{\label{normal-incidence-section}Normal incidence}

\begin{figure}
\begin{center} \includegraphics{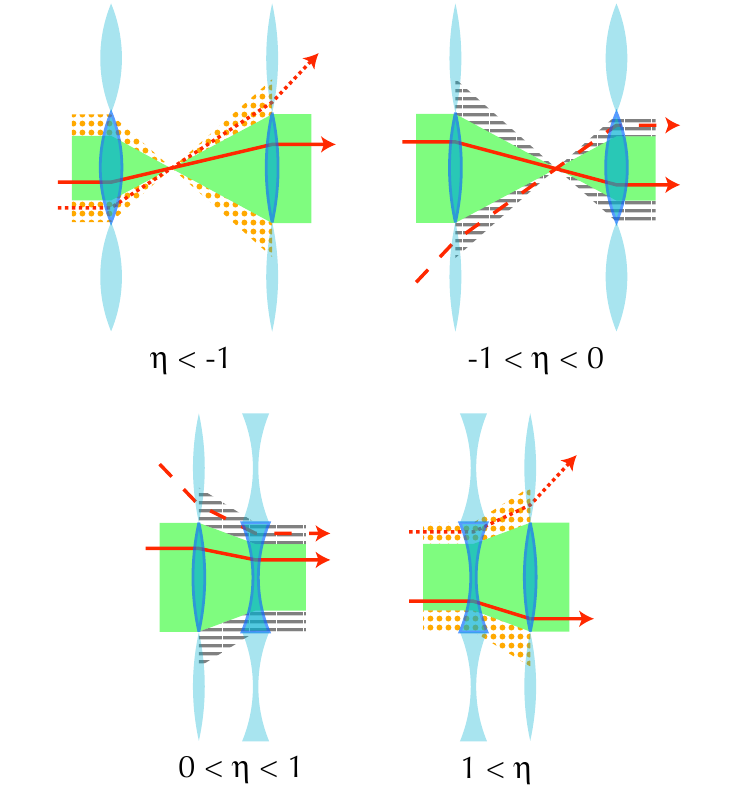} \end{center}
\caption{\label{normal-incidence-figure}Ray paths with angle of incidence $\alpha_1 = 0$ and angle of metarefraction $\alpha_2 = 0$, undergoing standard and non-standard metarefraction through CLA sheets.
The different diagrams represent the different $\eta$ ranges, namely
$\eta < -1$,
$-1 < \eta < 0$,
$0 < \eta < 1$, and
$1 < \eta$.
These $\eta$ ranges respectively correspond to the focal-length combinations \cite{Courtial-2008a}
$f_1 > 0$, $f_2 > 0$, $f_1 < f_2$ ($\eta < -1$);
$f_1 > 0$, $f_2 > 0$, $f_1 > f_2$ ($-1 < \eta < 0$);
$f_1 > 0$, $f_2 < 0$ ($0 < \eta < 1$); and
$f_1 < 0$, $f_2 > 0$ ($1 < \eta$).
Representative light rays for various kinds of metarefraction are shown:
those entering and exiting through corresponding lenslets (standard metarefraction) are shown as solid arrows;
those entering through the first lenslet with angle $\alpha_1 = 0$ but missing the corresponding second lenslet (non-standard metarefraction) are shown as dotted arrows;
and those exiting through the second lenslet with angle $\alpha_2 = 0$ but which have not passed through the corresponding first lenslet (a different kind of non-standard metarefraction) are shown as dashed arrows.
Light rays undergoing these three different kinds of metarefraction travel in different areas, respectively filled solid (standard metarefraction), dotted (wrong second lenslet), and striped (wrong first lenslet).
}
\end{figure}

The simplest case is arguably that of normal incidence and normal metarefraction, that is angle of incidence $\alpha_1 = 0$ and angle of metarefraction $\alpha_2 = 0$.
This is shown in Fig.\ \ref{normal-incidence-figure} for different $\eta$ ranges.
(Fig.\ \ref{normal-incidence-figure} also shows which combinations of focal lengths correspond to which $\eta$ range.)
Representative light rays undergoing standard metarefraction are shown as solid arrows.
They enter and exit the lenslets with angles $\alpha_1 = \alpha_2 = 0$.
Such light rays are restricted to the solid filled areas.

It is possible that light rays exit the second lenslet with $\alpha_2 = 0$, but that they did not enter through the corresponding first lenslet.
Here this is called \emph{non-standard metarefraction of the first kind}.
In Fig.\ \ref{normal-incidence-figure}, light rays undergoing non-standard metarefraction of the first kind are drawn as dashed arrows, and the area to which they are restricted is striped.

It is also possible that light rays enter through the first lenslet with angle $\alpha_1 = 0$ but miss the corresponding second lenslet, instead passing through a different lenslet in the second lenslet array and exiting the array with an angle $\alpha_2 \neq 0$.
This is called \emph{non-standard metarefraction of the second kind}.
Light rays undergoing non-standard metarefraction of the second kind, and the area to which they are restricted, are respectively shown as dotted arrows and dotted areas in Fig.~\ref{normal-incidence-figure}.

It is perhaps mildly surprising that non-standard metarefraction can occur already at normal incidence. whereby different kinds of non-standard metarefraction occur for different $\eta$ ranges.
The following sections investigate non-standard metarefraction for non-normal incidence and metarefraction.
They continue to use this section's fill-labelling scheme for standard metarefraction (solid) and non-standard metarefraction of the first (striped) and second (dotted) kind.

\subsection{\label{first-critical-angles-section}Onset of non-standard metarefraction of the first kind: the first critical angles}

Consider looking through a CLA sheet.
What you see in a specific direction is determined by the history of the light rays arriving at the eye from that direction.
When traced backwards from the eye (which is what ray-tracing software, such as POV-Ray \cite{POV-Ray}, does), light rays undergoing non-standard metarefraction of the first kind pass through a second lenslet and then miss the corresponding first lenslet.
In the previous section's labelling scheme, such light rays are dashed.

If such a dashed light ray is not absorbed, the direction in which the backwards-traced light ray leaves the CLA sheet (in terms of the direction in which the light ray actually travels, namely from the source to the eye, this is described by the angle of incidence) is different from the direction in which standard-metarefracted light rays leave the sheet.
On further backwards-tracing, the light ray usually hits an object (or point on an object) which is different from that which a standard-metarefracted light ray would have hit.
This ``wrong'' object (or point on an object) is then visible as a ghost image in the direction in which the light ray left the eye.
The fraction of light rays traced backwards from the eye position that have passed through a given second lenslet and that miss the corresponding first lenslet determines the brightness of the ghost image seen in the direction of the second lenslet.

If such a dashed light ray is absorbed, then the trajectory of the actual light ray would end on the absorber and never reach the eye.
If the light ray was nevertheless traced backwards from the eye in the direction from which the light ray would have arrived had it not been absorbed, then it would also end on the absorber.
In that direction, the observer would therefore see the colour of the absorber: black.
In this case, the fraction of all the light hitting the eye from an entire second lenslet that is absorbed determines the factor by which the intensity of the standard-refracted image is dimmed.

\begin{figure}
\begin{center}
\includegraphics{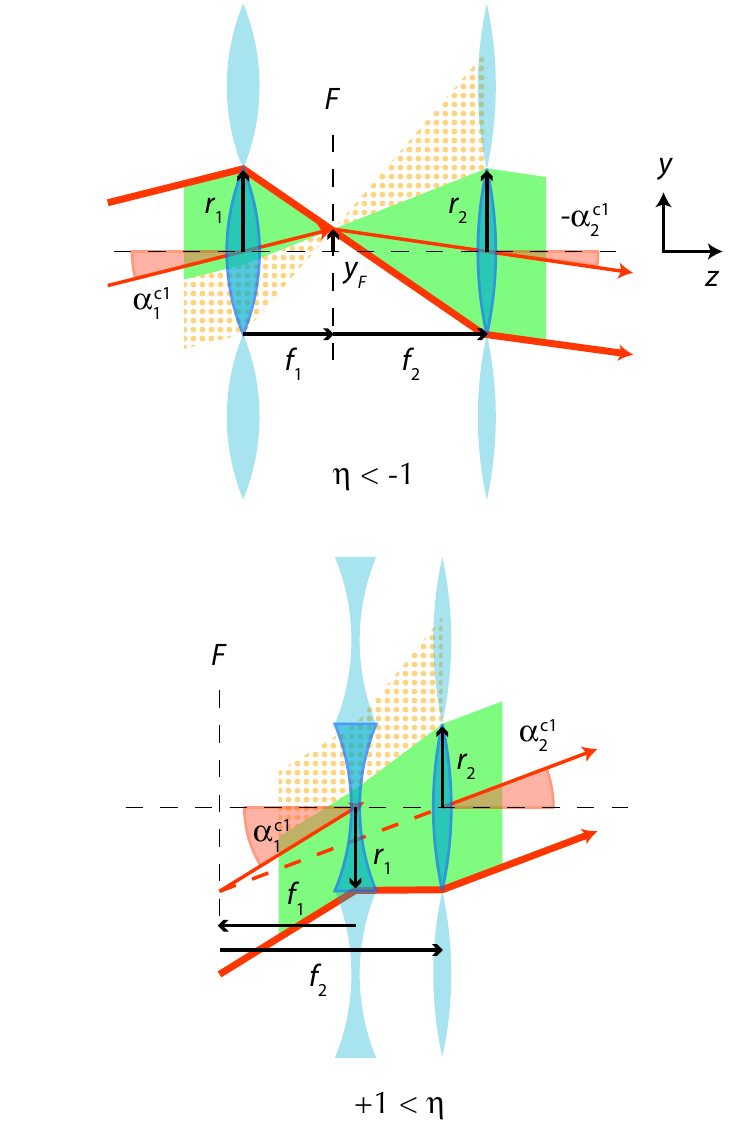}
\end{center}
\caption{\label{first-critical-angles-figure}Calculation of the first critical angles which mark the onset of ghost-image metarefraction.
The thick arrows indicate the ``critical'' light ray used for the calculation of the critical angles.
$\alpha_1^{c1}$ and $\alpha_2^{c1}$ are the first critical angles of incidence and metarefraction, respectively;
$f_1$ and $f_2$ are the focal lengths of the left and right lenslets, respectively;
$r_1$ and $r_2$ is the radius of the first and second lenslet, respectively (whereby a lenslet's aperture radius has the same sign as its focal length -- see main text);
$y_F$ is the $y$-coordinate (measured from the optical axis) of the point at which all light rays intersect the focal plane, $F$, at the first critical angle of incidence.}
\end{figure}

Non-standard metarefraction of the first kind occurs over a specific range of angles of incidence and metarefraction.
Fig.\ \ref{first-critical-angles-figure} sketches light rays passing through a CLA sheet with the greatest positive incidence angle for which no non-standard metarefraction of the first kind occurs: no dashed light rays are present, but the slightest increase in the angle of incidence would mean that the solid rays no longer completely cover the right lenslet and that the gap is filled with dashed rays (using the labelling scheme introduced in the previous section).
This angle of incidence is called the \emph{first critical angle of incidence}, $\alpha_1^{c1}$.
No first-kind non-standard metarefraction occurs for any angles of incidence $\alpha_1$ with a modulus less than $\alpha_1^{c1}$, that is for
\begin{equation}
|\alpha_1| < \alpha_1^{c1}.
\label{alpha1c1-definition}
\end{equation}
The corresponding range of angles of metarefraction, $\alpha_2$ is then
\begin{equation}
|\alpha_2| < \alpha_2^{c1},
\label{alhpa2c1-definition}
\end{equation}
where $\alpha_2^{c1}$ is the \emph{first critical angle of metarefraction}, which is related to the first critical angle of incidence by the equation
\begin{equation}
\tan \alpha_1^{c1} = |\eta| \tan \alpha_2^{c1}.
\label{first-critical-angles-relation}
\end{equation}
This last equation is, of course, equation (\ref{tan-equation}) with $\eta$ replaced by $|\eta|$ to ensure that $\alpha_1^{c1}$ and $\alpha_2^{c1}$ have the same sign.

Fig.\ \ref{first-critical-angles-figure} actually shows separate diagrams, one for the case $\eta < -1$, the other for $+1 < \eta$.
(There is no diagram for $-1 < \eta < +1$, as in this range ghost images occur even at normal incidence -- see section \ref{normal-incidence-section}.)
$\alpha_1^{c1}$ can be calculated for both cases at the same time as follows.
Define the aperture radius of a lenslet to be positive if its focal length is positive; similarly, a lenslet's aperture radius is negative if its focal length is negative.
(Fig.\ \ref{first-critical-angles-figure}, just like Figs \ref{normal-incidence-figure}, \ref{second-critical-angles-figure} and \ref{third-critical-angles-figure}, is drawn for the simplest case, $|r_1| = |r_2|$.
Nevertheless, in the range $1 < |\eta|$ the calculations of the critical angles hold also if this is not the case.)
In the coordinate system indicated in Fig.\ \ref{first-critical-angles-figure}, the light ray that passes through the first lenslet with the first critical angle of incidence and at vertical coordinate $y=r_1$ passes through the second lenslet at $y=-r_2$.
(Note that $r_1$ is negative in the case $+1 < \eta$.)
This ``critical'' light ray is marked in both diagrams in Fig.\ \ref{first-critical-angles-figure}. 

All light rays that enter the first lenslet with the first critical angle of incidence intersect in the same point in the focal plane, at $y$-coordinate $y_F$.
In the following, the critical light ray will be used to calculate $y_F$.
Then the principal light ray through the center of the first lenslet, which passes straight through the lenslet, will be used to calculate $\alpha_1^{c1}$.

Between the lenslets, the slope of the critical light ray is $-(r_1 + r_2) / (f_1 + f_2)$.
In particular, this is the slope of the critical light ray between the first lenslet and the focal plane, which implies that
\begin{equation}
\frac{y_F - r_1}{f_1} = - \frac{r_1 + r_2}{f_1 + f_2},
\end{equation}
so
\begin{equation}
y_F = r_1 - \frac{r_1 + r_2}{f_1 + f_2} f_1,
\label{y_F-equation}
\end{equation}
From the principal light ray that passes through the first lenslet's center with the first critical angle of incidence it can be seen that
\begin{equation}
\tan \alpha_1^{c1} = \frac{y_F}{f_1}.
\end{equation}
Solving for $\alpha_1^{c1}$ and substituting the expression for $y_F$ in equation (\ref{y_F-equation}), this becomes
\begin{equation}
\alpha_1^{c1} = \tan^{-1} \left( \frac{r_1}{f_1} - \frac{r_1 + r_2}{f_1 + f_2} \right).
\label{alpha1c1-equation}
\end{equation}
This is the expression for the first critical angle, valid in the range $1 < |\eta|$.

It is useful to try out equation (\ref{alpha1c1-equation}) for the simplest case (and for which the figures are drawn), namely $|r_1| = |r_2| = r$.
For a CLA sheet's $\eta$ value to fall into the range $\eta < -1$, its focal lengths have to satisfy the inequalities $f_1 > 0$, $f_2 > 0$, and $f_1 < f_2$ (Fig.\ \ref{normal-incidence-figure}).
The fact that both focal lengths are positive means, according to our sign convention for aperture radii, that both aperture radii are also positive, so $r_1 = r_2 = r$.
Equation (\ref{alpha1c1-equation}) then becomes
\begin{eqnarray}
\alpha_1^{c1} = \tan^{-1} \left( \frac{r}{f_1} - \frac{2 r}{f_1 + f_2} \right) = \tan^{-1} \frac{r (f_2 - f_1)}{f_1 (f_1 + f_2)}. \nonumber \\
\label{alpha1c1-negative-eta-equation}
\end{eqnarray}
As $f_1 < f_2$ and the values of all parameters in the argument of the inverse tangent are positive, the critical angle $\alpha_1^{c1}$ is also positive.
According to the definition of $\alpha_1^{c1}$, equation (\ref{alpha1c1-definition}), this means there is a range of incidence angles $\alpha_1$, centered around normal incidence, for which no non-standard metarefraction of the first kind occurs, which is what was expected from the discussion in section \ref{normal-incidence-section} and in this section so far.

In the case $+1 < \eta$, a CLA sheet's focal lengths have to satisfy $f_1 < 0$ and $f_2 > 0$ according to Fig.\ \ref{normal-incidence-figure}.
According to our sign convention, the aperture radii are then $-r_1 = r_2 = r$, so now equation (\ref{alpha1c1-equation}) becomes
\begin{equation}
\alpha_1^{c1} = \tan^{-1} \frac{-r}{f_1}.
\end{equation}
Reassuringly, as $f_1 < 0$, $\alpha_1^{c1}$ is again positive.



\subsection{\label{second-critical-angles-section}Onset of non-standard metarefraction of the second kind: the second critical angles}

Whenever non-standard metarefraction of the second kind occurs, not all the light that enters through the first lenslet subsequently passes through the corresponding second lenslet.
The standard-refracted image is therefore dimmed.
Light that misses the corresponding second lenslet either leads to ghost images at another angle of metarefraction, or (if filtered out with absorbers) it is absorbed entirely.


\begin{figure}
\begin{center}
\includegraphics{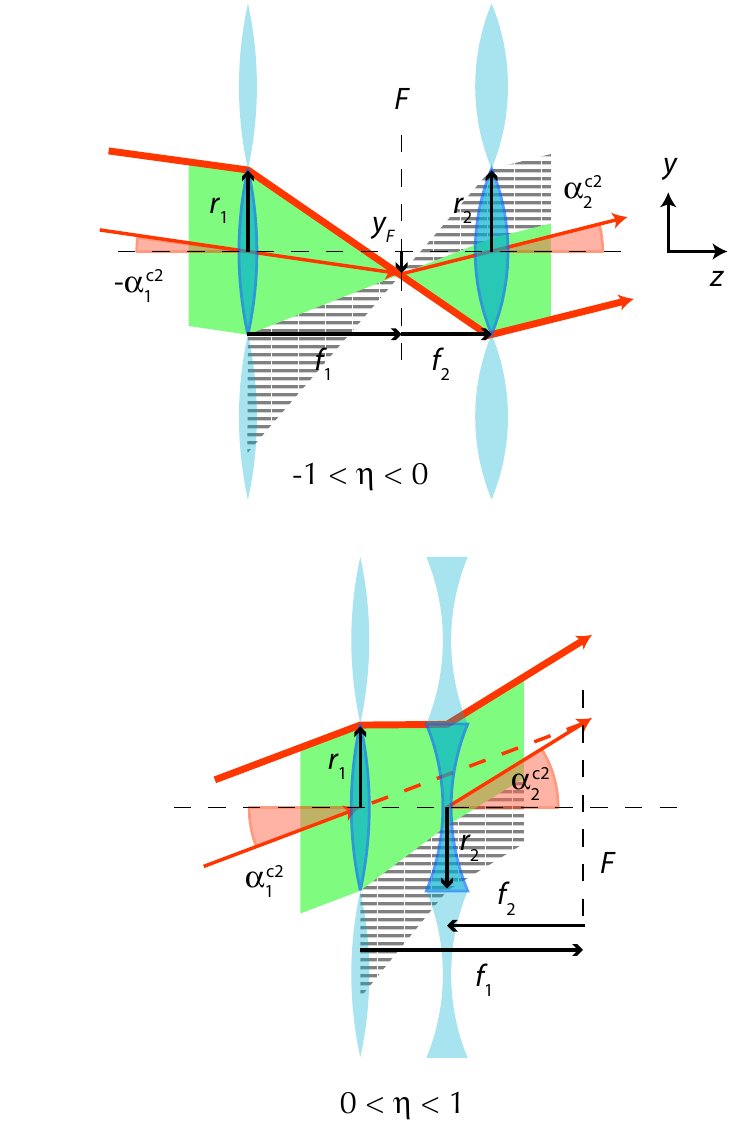}
\end{center}
\caption{\label{second-critical-angles-figure}Calculation of the second critical angles, which mark the onset of dimming of standard-refracted light.}
\end{figure}

In analogy to the first critical angles, the second critical angles of incidence and metarefraction, $\alpha_1^{c2}$ and $\alpha_2^{c2}$, are respectively defined as the modulus of the angle of incidence and metarefraction at which the dimming of the standard-refracted image starts to occur.
Fig.\ \ref{second-critical-angles-figure} shows diagrams of light passing through confocal lenslet arrays at the second critical angles of incidence and metarefraction, drawn for the cases $-1 < \eta < 0$ and $0 < \eta < 1$.
(For $\eta < -1$ and $+1 < \eta$, dimming occurs even at normal incidence.)

In a similar way to the derivation of the first critical angle of incidence, the second critical angle of metarefraction can be shown to be
\begin{equation}
\alpha_2^{c2} = \tan^{-1} \left( \frac{r_2}{f_2} - \frac{r_1 + r_2}{f_1 + f_2} \right),
\label{alpha2c2-equation}
\end{equation}
from which the second critical angles of incidence can be calculated through the analog of equation (\ref{first-critical-angles-relation}),
\begin{equation}
\tan \alpha_1^{c2} = |\eta| \tan \alpha_2^{c2}.
\label{second-critical-angles-relation}
\end{equation}
The expressions for the second critical angles, equations (\ref{alpha2c2-equation}) and (\ref{second-critical-angles-relation}), are valid for $|\eta| < 1$.



\subsection{\label{third-critical-angles-section}Disappearance of standard metarefraction: the third critical angles}

\begin{figure}
\begin{center}
\includegraphics{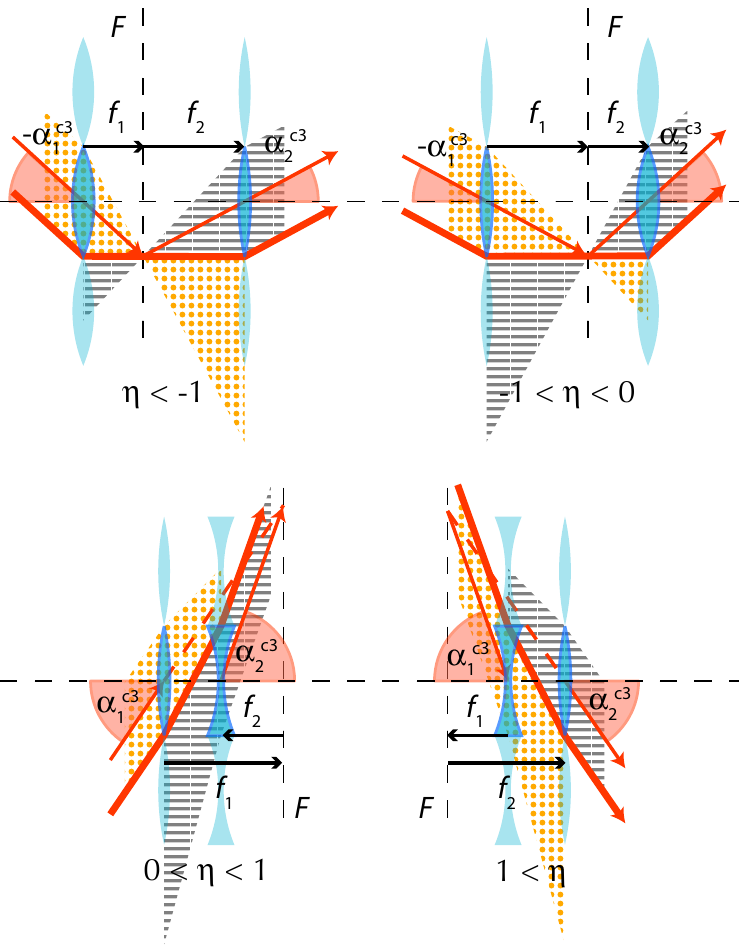}
\end{center}
\caption{\label{third-critical-angles-figure}Calculation of the third critical angles.
Like in previous figures, the critical light ray is indicated as a thick solid arrow.}
\end{figure}

Under the assumptions for which the results presented in this section are derived, at least some normally-incident light is standard-refracted:
in the range $|\eta| < 1$, all the normally incident light is standard-refracted;
in the range $|\eta| > 1$, some of the light exits through the ``wrong'' lenslet and forms ghost images at higher angles of metarefraction, but at least part of the light passes through the corresponding second lenslet and is therefore normally refracted.
As the angle of incidence is increased beyond $\alpha_1^{c1}$, the region of standard metarefraction shrinks and then disappears completely.
The angles of incidence and metarefraction for which standard metarefraction disappears are called the third critical angles, $\alpha_1^{c3}$ and $\alpha_2^{c3}$; this case is shown in Fig.~\ref{third-critical-angles-figure}.

The calculation of the third critical angles starts with the observation that in all $\eta$ ranges both third critical angles have to be positive as otherwise there would be no standard metarefraction at normal incidence.
This means that 
the third critical angles are simply the absolute values of the angles of incidence and metarefraction sketched in Fig.\ \ref{third-critical-angles-figure}.

As before, the slope of the critical light ray (see Fig.\ \ref{third-critical-angles-figure}) between lenslets can be calculated.  This slope is
\begin{equation}
\frac{r_1 - r_2}{f_1 + f_2}.
\end{equation}
From this it is possible to calculate the $y$-coordinate of the point at which the critical light ray -- and indeed all light rays with the same angle of incidence -- intersect.
Placing the optical axis at $y=0$, as before, the $y$-coordinate of this intersection point is
\begin{equation}
y_F =  -r_1 + f_1 \frac{r_1 - r_2}{f_1 + f_2}.
\end{equation}
This means the corresponding angle of incidence is given by the equation
\begin{equation}
\tan \alpha_1 = \frac{y_F}{f_1} = - \frac{r_1}{f_1} + \frac{r_1 - r_2}{f_1 + f_2}.
\end{equation}
The third critical angle of incidence is the modulus of the angle of incidence, which is therefore
\begin{equation}
\alpha_1^{c3} = \left| \tan^{-1} \left( - \frac{r_1}{f_1} + \frac{r_1 - r_2}{f_1 + f_2} \right) \right|.
\label{alpha1c3-equation}
\end{equation}

The third critical angle of metarefraction can be calculated similarly.
The result is
\begin{equation}
\alpha_2^{c3} = \left| \tan^{-1} \left( \frac{r_2}{f_2} + \frac{r_1 - r_2}{f_1 + f_2} \right) \right|.
\label{alpha2c3-equation}
\end{equation}

Equations (\ref{alpha1c3-equation}) and (\ref{alpha2c3-equation}) are valid for any value of~$\eta$.

\section{Comparison with simulations}

\begin{figure}
\begin{center} \includegraphics{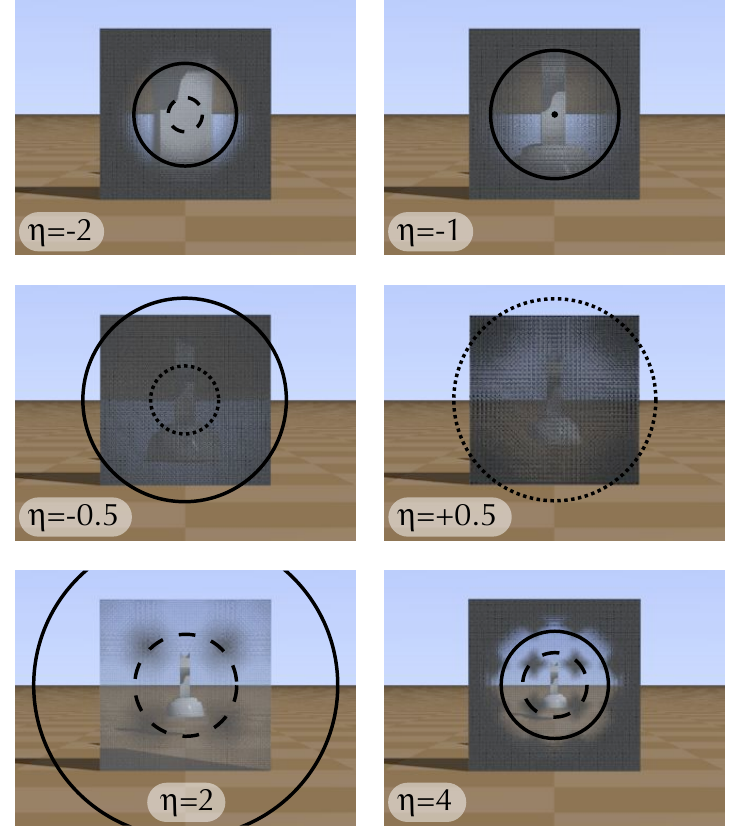} \end{center}
\caption{\label{etas-with-critical-angles-figure}Comparison of calculated critical angles with simulations.
Circles on the sheet indicate a constant angle of metarefraction, whereas the first, second, and third critical angles of metarefraction (see Table \ref{critical-angles-table}) are respectively indicated by dashed, dotted, and solid circles.
Note that the circles corresponding to the first and second angles of metarefraction in the case $\eta = -1$ have radius zero and are therefore shrunk to a point in the sheet center.
Note also that the third critical angle in the case $\eta = +0.5$ is too large to be shown.
In units of the side length of each floor tile, the side length of the sheet is 1 and the camera was positioned a distance $z=6$ in front of the sheet.
As each sheet's center corresponds to a $0^\circ$ angle of metarefraction, the angle of metarefraction in the middle of each sheet edge is therefore $4.8^\circ$, that at the corners of each sheet is $6.7^\circ$.
The frames were simulated like those in Fig.\ \ref{etas-figure}, but with the following differences: 
Fig.\ \ref{etas-figure} was simulated for CLA sheets consisting of lenslets with square apertures and a short depth of focus, with the simulated camera focussed onto the image of the chess piece;
here the lenslet apertures were circular and the depth of focus infinite.}
\end{figure}

Fig.\ \ref{etas-with-critical-angles-figure} shows the view through CLA sheets with the same values of $\eta$ as those shown in Fig.\ \ref{etas-figure}, but with a few changes that allow more direct comparison with the results from the previous sections for the calculation of the critical angles.
Because Fig.\ \ref{etas-with-critical-angles-figure} was simulated with the camera placed on the sheet normal through the center of each CLA sheet, the center of each sheet corresponds to an angle of metarefraction $\alpha_2 = 0^\circ$.
Light rays from a point a distance $r$ from the sheet center correspond to an angle of metarefraction
\begin{equation}
\tan \alpha_2 = \frac{r}{z},
\end{equation}
where $z$ is the distance of the camera from the sheet.
This means that the points on the CLA sheet that are seen in light that has left the CLA sheet with the same angle of metarefraction lie on circles around the sheet center.

\begin{table} 
\begin{center}
\begin{tabular}{c|cccccc}
$\eta$ & -2 & -1 & -0.5 & +0.5 & 2 & 4 \\
\hline
$f_1$ & 0.025 & 0.04 & 0.05 & 0.05 & -0.025 & -0.02 \\
$f_2$ & 0.05 & 0.04 & 0.025 & -0.025 & 0.05 & 0.08 \\
$\alpha_2^{c1}$ & $0.95^\circ$  & $0^\circ$ & N/A & N/A & $2.9^\circ$ & $1.8^\circ$ \\
$\alpha_2^{c2}$ & N/A & $0^\circ$ & $1.9^\circ$ & $5.7^\circ$ & N/A & N/A \\
$\alpha_2^{c3}$ & $2.9^\circ$ & $3.6^\circ$ & $5.7^\circ$ & $17^\circ$ & $8.5^\circ$ & $3.0^\circ$
\end{tabular}
\end{center}
\caption{\label{critical-angles-table}Table of critical angles of metarefraction, calculated for the CLA sheets shown in Fig.\ \ref{etas-with-critical-angles-figure}.
The cases for which there are no first or second critical angles (as the first or second kind of non-standard metarefraction occurs for all angles, so there is no onset) are marked ``N/A''.
In all cases, $|r_1| = |r_2| = 0.0025$; the sign of $r_1$ is that of $f_1$, the sign of $r_2$ is that of $f_2$.}
\end{table}

Table \ref{critical-angles-table} lists the parameters for which the simulations in Fig.\ \ref{etas-with-critical-angles-figure} (and those in Fig.\ \ref{etas-figure}) were performed, and the corresponding critical angles of metarefraction calculated from these parameters.
The circles corresponding to these critical angles of metarefraction are superposed on the simulations in Fig.\ \ref{etas-with-critical-angles-figure}.

Perhaps the most obvious feature of Fig.\ \ref{etas-with-critical-angles-figure} is that outside the (solid) circle corresponding to the third critical angle of metarefraction the intensity falls of very rapidly and very little light passes the sheet ouside this circle and reaches the camera.
This confirms the considerations in section \ref{third-critical-angles-section}, according to which no light that has passed the sheet outside this circle should reach the camera (Fig.\ \ref{etas-with-critical-angles-figure} was calculated with absorbers that remove all light rays undergoing non-standard metarefraction), but only approximately:
there is clearly some light that passes through the sheet just outside the solid circle and then reaches the camera, most notably in the case $\eta = 4$.
The origin for this ``leakage'' is not completely clear, but one possible reason could be the fact that in the simulations for Fig.\ \ref{etas-with-critical-angles-figure}, the apertures of corresponding lenslets were not separated exactly by the sum of the lenslets' focal lengths.
This is due to the fact that corresponding lenslets are set up such that the centers of their outside surfaces were separated by the sum of the lenslets' focal lengths, but because the lenslet surfaces are curved, the edges of corresponding lenslets -- the effective apertures -- were separated by a slightly different distance.
In the simulations, this separation is within less than 1\% of the lens-center separation.

The light dimming due to non-standard metarefraction of the second kind is not represented at all in our ray-tracing simulations.
This can be understood by the following argument which considers the light from an arbitrary small part of the chess piece's surface that subsequently passes through a specific lenslet in the first lenslet array.
Physically, for given lighting conditions the power of this light is fixed; losing any of the power leads to dimming.
Exactly such loss of power happens for angles above the second critical angles, as represented by the dotted light rays in the cases $\eta < -1$ and $1 < \eta$ in Fig.\ \ref{normal-incidence-figure}:  such light rays get refracted into the wrong angle of metarefraction, and are lost from the correct angle of metarefraction (in the case of Fig.\ \ref{normal-incidence-figure}, $0^\circ$).
However, if the solid light rays are traced backwards, none of the light appears lost in any way, so this particular dimming is consequently not represented.
(Perhaps it is helpful to look at this in the following, slightly different, way.
The solid area in Fig.\ \ref{normal-incidence-figure} can be seen as the standard-refracted \emph{beam}.
In those cases where dotted light rays occur, namely $\eta < -1$ and $1 < \eta$, the beam diameter is smaller as it enters the first lenslet compared to when it leaves the second lenslet.
This means that the power contained in the beam gets spread across a larger area, which should lead to a reduction in intensity.)

The effect of non-standard metarefraction of the first kind \emph{is} represented in Fig.\ \ref{etas-with-critical-angles-figure}, but its effect is far less obvious than that of the third critical angles.
The effect of the lenslet arrays' square symmetry obscures the effect further, for example in the frames corresponding to $\eta = 2$ and $\eta = 4$. 
Nevertheless, the dashed circle (which represent the first critical angles of metarefraction) arguably gives an indication of the size of the clear circle in the center of each frames, most notably for $\eta = 2$ and $\eta = 4$.


\section{\label{conclusions-section}Conclusions} 

This paper starts to study optical imperfections of generalized CLA sheets, specifically light passing through non-corresponding lenslets, which leads to non-standard metarefraction.



The formulae that were derived for the critical angles contain only the focal lengths of the lenslets and the radius of each lenslet.
They therefore provide a clear guide on what needs to be done to increase the field of view.
However, it should be noted that an increase in the radius of each lenslet aperture without a corresponding increase in the focal length increases the angle at which light rays travel, and with it aberrations, unless great care is taken in the lens design.
Therefore the field of view needs to be traded off against imaging quality.

Several questions remain.
In this paper only the simplest generalized CLAs were studied, so it is natural to examine the critical angles for more complex, generalized, CLAs \cite{Hamilton-Courtial-2009b}.
Similarly, little is known about the field of view of the closely related Dove-prism sheets \cite{Hamilton-Courtial-2008a}, and of combinations of Dove-prism sheets that perform negative metarefraction \cite{Courtial-Nelson-2008} and light-ray rotation~\cite{Hamilton-et-al-2009}.





\section*{Acknowledgments}

JC is a Royal Society University Research Fellow.


\end{document}